# Profile of James Peebles, Michel Mayor, and Didier Queloz: 2019 Nobel Laureates in Physics

Neta A. Bahcall and Adam Burrows,
Department of Astrophysical Sciences,
Princeton University, Princeton, NJ 08544

Mankind has long been fascinated by the mysteries of our Universe: How old and how big is the Universe? How did the Universe begin and how is it evolving? What is the composition of the Universe and the nature of its dark-matter and dark-energy? What is our Earth's place in the cosmos and are there other planets (and life) around other stars?

The 2019 Nobel Prize in Physics honors three pioneering scientists for their fundamental contributions to these cosmic questions – Professors James Peebles (Princeton University), Michel Mayor (University of Geneva), and Didier Queloz (University of Geneva and the University of Cambridge) – "for contributions to our understanding of the evolution of the universe and Earth's place in the cosmos," with one half to James Peebles "for theoretical discoveries in physical cosmology," the other half jointly to Michel Mayor and Didier Queloz "for the discovery of an exoplanet orbiting a solar-type star."

### 'The Evolution of the Universe'

James Peebles was born in Southern Manitoba, Canada, in 1935; he moved to Princeton University for graduate-school in 1958 and remained in Princeton as the Albert Einstein Professor of Physics (now emeritus) for his entire career. He has long been considered one of the fathers of modern cosmology, the science of the origin and evolution of the Universe. As the Nobel committee states "James Peebles took the cosmos, with its billions of galaxies and galaxy clusters. His theoretical framework, developed over two decades, is the foundation of our modern understanding of the Universe's history, from the Big-Bang to the present day." The spectacular advances in cosmology over the past half a century have been driven by a combination of both observations and theory. The revelation that our Universe has begun from a hot and dense Big-Bang 13.8 billion years ago and has been expanding and evolving ever since, under the influence of the still mysterious dark matter and dark energy that dominate the cosmos, in addition to the ~5% baryons that makes up the stars, planets, and all of us – is a triumph for both observational and theoretical science.

The science of cosmology started in earnest mostly with Einstein's theory of general relativity. Alexander Friedman, Willem de Sitter, and Georges Lemaître, using Einstein's equations, recognized that our Universe is not static - it must be either expanding or contracting and Hubble's remarkable observational discovery of the expanding Universe in 1929 confirmed this. The observed expansion implied that the Universe must have started from a compact state, but this idea took a long time to be accepted. In the 1940s, George Gamow, Ralph Alpher, and Robert Herman investigated the consequences of a hot Big Bang – a Universe that started from a small, very dense, and very hot stage – and made predictions that could be tested. They calculated that light elements should have been formed in the hot Big Bang and that a huge amount of radiation, the remnant of which (the Cosmic Microwave Background, or CMB) should still fill the entire Universe today, must have been generated.

However, when Peebles began working in cosmology in the 1960's, at the invitation of his Princeton mentor and friend Professor Robert Dicke, the field was vague, not well quantified, and had only modest experimental support. In 1965, Peebles, with his colleagues Dicke, Wilkinson, and Roll, recalculated the black-body temperature of the remnant radiation expected from the aforementioned hot Big Bang and estimated it to be ~10 K today and Dicke and collaborators began to look for it [1]. That same year the signal of the CMB was indeed detected by Arno Penzias and Robert Wilson at Bell Labs in Holmdel New Jersey [2]. Penzias and Wilson conferred with Peebles and Dicke to understand the observed signal, which corresponded to a radiation temperature of 3.5 K. The discovery of the CMB radiation confirmed the hot Big Bang model and opened the door to a greatly improved understanding of physical cosmology.[*] Penzias and Wilson were awarded the 1978 Nobel prize for their observational discovery of the CMB.

With the discovery of the CMB, Peebles investigated its impact on the formation of nuclei in the first few minutes of the hot dense Universe – the so-called Big-Bang Nucleosynthesis (BBN). While BBN had been worked on previously, the influence of the hot CMB has not been considered in detail. Peebles used the measured temperature of the CMB, extrapolated to the hot temperature during the early Universe and calculated the resulting nucleosynthesis. From his calculations of the hot Big-Bang, Peebles determined the abundance of light elements created in the early Universe. His estimate for primordial Helium of 27-30% of all nuclei by mass is similar to the currently observed primordial Helium abundance (25%).

Peebles continued his exploration of physical cosmology in numerous areas. He predicted the temperature fluctuations in the spectrum of the CMB – the primordial seeds of structure formation in the Universe; it took many years of continuously improved experimental techniques to finally detect the CMB fluctuations (first with COBE [the Cosmic Background Explorer mission], and later with CMB measurements by BOOMERanG [the Ballon Observations of Millimetric Extragalactic Radiation And Geophysics experiment], WMAP [the Wilkinson Microwave Anisotropy Map], and the Planck mission). He helped provide theoretical evidence for dark matter in spiral galaxies by showing that they would not be stable at their observed flattened shape without a more stabilizing gravitational potential of a massive halo to surround them. Peebles pioneered the cosmological model of 'Cold Dark Matter,' which is the best current model for dark matter, based on evidence from observations of large-scale structure. He was a pioneer in the quantitative study of the observed distribution of galaxies, using the powerful statistical tools of the two-point and higher-order correlation functions, and applied these tools to large surveys of galaxies. He pioneered models for cosmic structure formation and evolution through the gravitational instability picture in an expanding Universe. Moreover, Peebles has also authored several outstanding books in cosmology, used by generations of students and scientists at all levels: "Physical Cosmology" [3], "Large Scale Structure of the Universe" [4], "Principles of Physical Cosmology" [5], "Finding the Big Bang" [6], and his most recent book just being completed "Cosmology's Century" [7].

"Just about every advance in our understanding of cosmology has been boosted by the work of Jim Peebles," says Princeton cosmologist Jeremiah Ostriker. "Big bang nucleosynthesis, the growth of cosmic structure, the existence of dark matter, and so many other advances in our understanding were shepherded by Jim Peebles' work and wisdom." Thanks to his efforts, cosmology is now a solid, robust scientific field, with both theory and observation going hand-in-hand to reveal the amazing and evolving Universe. While some fundamental questions remain open –including what is the dark matter and what is the nature of the mysterious dark energy - cosmologists can now explain, quantify, and make future predictions for the evolution of the Universe.

"Jim is one of the true giants in the field," wrote Paul Steinhardt, Peebles' colleague at Princeton. "His work transformed our understanding of the hot, expanding Universe from qualitative to precise, revealed the existence of dark matter, and pointed out the puzzles that remain."

**'Earth's Place in the Cosmos'**

Before 1995, a few astronomers were trying to develop spectroscopic techniques with which to measure the Doppler wobble of a parent star due to the possible presence of planetary companions. Some considered this indirect method more viable than direct imaging, since the dim planet would be difficult to detect from under the glare of the parent star. Another indirect technique being pursued was astrometry, whereby the reflex motion of the star due to the gravitational tug of an orbiting planetary companion would signal the planet's presence. A Jupiter-mass planet at the distance that Jupiter is from the Sun, but orbiting a nearby solar-mass star, would induce a stellar wobble of ~12 m s$^{-1}$ (only a bit faster than a human can run) in the stellar spectrum and an astrometric displacement of ~0.005 astronomical units (AU). One AU is the distance between the Earth and the Sun. The latter at a distance from the solar system of ~10 parsecs is equivalent to an angular displacement of the nearby star of ~0.5 milliarcsecs. However, even now, almost twenty-five years since 1995, such astrometric detection techniques have yet to prove themselves.

This left the Doppler technique. However, astronomers were routinely achieving only ~200 m s$^{-1}$. Some astronomers, using hydrogen fluoride or iodine internal spectral calibrators and good thermal control, had achieved ~20 m s$^{-1}$, but had discovered nothing. In fact, few astronomers were working on this problem and without discoveries this field had all the earmarks of an astronomical backwater. Stellar astronomy was not attracting new blood, the hot topics were in extragalactic astronomy and cosmology, and a few false or dubious detections had tainted the enterprise. The center-of-gravity of exoplanet research then resided with the search for extraterrestrial intelligence ("SETI"), a program that has yet to achieve true scientific legitimacy and claims no discoveries. The only bright spot was the discovery by Alex Wolszczan in 1992 of three putative planets orbiting the radio pulsar PSR 1257+12, exploiting the exquisite timing possible in radio pulsar research. However, since then few other such systems have been unambiguously discovered and this system itself has not been understood nor followed up fruitfully by other techniques. The discovery did not give birth to a vigorous and robust new field; and PSR 1257+12 remains an unexplained oddity that has proven a scientific cul-de-sac.

In this environment and context, Michel Mayor and Didier Queloz of Geneva Observatory had developed and were fielding a precision Doppler technique and in 1995 were regularly monitoring a list of nearby stars. The expectation was that whatever detections were to be made would be of planets in long-period orbits of years. All groups were looking for "clones" of the solar system, and a theorist had earlier that year predicted that giant planets would be found at orbital separations of more than 4.5 AU and, therefore, with long orbital periods. Jupiter's orbital period is ~12 years. Moreover, the Doppler shifts signaling the possible presence of such a planet would evolve slowly. Hence, there was no perceived hurry.

However, in the second half of 1995 all this changed. Mayor and Queloz detected something strange. But could a Jupiter-mass object in a ~0.05-AU orbit around a sun-like star survive under such intense irradiation? Could such a planet persist for a Hubble time? They sought out the help of a theorist who was working on basic giant planet theory, who quickly performed some simple calculations of giant planet evolution, and found that its gravitational potential well could indeed keep most of its mass bound for billions of years. Armed with this information, in October of 1995 at a meeting in Florence, Italy Mayor and Queloz made their announcement. Using their state-of-the-art method, they had detected an exoplanet, 51 Peg b, with a minimum mass of ~0.5 Jupiter masses, but in a surprisingly tight ~4.5-day orbit of ~0.05 AU, one hundred times closer to its star than Jupiter is to the Sun.[**]

However, though the audience at Florence was intrigued, many in the room did not believe the result. A giant planet found where it was not expected, at an uncomfortable orbital distance, discovered by a technique that had never before produced reliable astronomy at the needed level of precision, was problematic. Could the result be an artifact of star spots? Could the team be seeing a more prosaic binary star system, but almost face-on? The initial reaction was the traditional one of prudence and skepticism in the face of a radical discovery. Their publication [9] was refereed by three scientists - two accepted the paper and one rejected it, but the majority prevailed. It would take more than a year before the authenticity of this find and its implications for planetary science were accepted and the field has never looked back.

To date, we know of more than 4000 exoplanets, many discovered by the Kepler space telescope using another technique (the "transit" method, meauring the tiny decrease in star-light when the planet transits in front of the star ), most are "super-Earths" and "sub-Neptunes," and many are in exotic multiple planet systems executing unexpected orbits − we have learned that our solar system is by no means typical. The Hubble and Spitzer space telescopes have now directly detected the light from scores of transiting giant exoplanets, inaugurating the remote sensing era of exoplanet studies. Planets in very wide orbits have been discovered by direct, high-contrast imaging around nearby stars. It was with this technique that many had thought exoplanets would first be discovered. Metaphorically, astronomers had been looking to little effect under "lampposts," but Mayor and Queloz, in part by luck, but enabled by technical advance and dogged determination, showed the way, thereby creating a new astronomy and a new planetary science.

Currently, a large and increasing fraction of astronomers is engaged in exoplanet research. It is one of the fastest growing segments of international astronomy. Furthermore, many young astronomers are now embarking upon careers in exoplanetary science. Space missions are being designed to discover, probe, and characterize exoplanets. People have redirected their careers to get involved. Exoplanet research has emerged as one central focus of astronomy's future. There is much excitement in the air that shows no signs of abating. This is Mayor and Queloz' legacy.

*Penzias and Wilson decided to publish their results simultaneously with Dicke and Peebles; In the first paper, Dicke, Peebles, Roll, and Wilkinson[1] showed the importance of the CMB as evidence for the hot Big Bang model. In the second paper, by Penzias and Wilson titled "A Measurement of Excess Antenna Temperature at 4080 Megacycles per Second"[2], they reported the existence of a 3.5 K residual background and attributed it to a "possible explanation" as that given by Dicke et al. in their companion letter[1].

** At the same meeting, in part motivated by the impending announcement of 51 Peg b, a Caltech group[8] announced the detection of the first unimpeachable brown dwarf, Gliese 229B. October of 1995 was a rather exciting month for many of those concerned.